\documentclass[amssymb,prb,twocolumn,floats,amsmath,showpacs]{revtex4}
\usepackage[T1]{fontenc}
\usepackage{bm}
\usepackage{graphicx}% Include figure files
\usepackage{amssymb}
\usepackage{amsfonts}
\usepackage{amsmath}
\usepackage{epstopdf}
\usepackage{color}

\begin{document}

\title{\textit{Single-color}, \textit{in situ} photolithography marking of individual CdTe/ZnTe Quantum Dots containing a single Mn$^{2+}$ ion}

\author{K.~Sawicki}
\author{F.~K.~Malinowski}
\author{K.~Ga\l{}kowski}
\author{T.~Jakubczyk}
\author{P.~Kossacki}
\author{W.~Pacuski}
\author{J.~Suf\mbox{}fczy\'{n}ski}
\email{Jan.Suffczynski@fuw.edu.pl}
\affiliation{Institute of Experimental Physics, Faculty of Physics, University of Warsaw, Pasteura 5 St., PL-02-093 Warsaw, Poland}

\date{\today}

%################################################################# ABSTRACT

\begin{abstract}
A simple, \textit{single-color} method for permanent marking of the position of individual self-assembled semiconductor Quantum Dots (QDs) at cryogenic temperatures is reported. The method combines \textit{in situ} photolithography with standard micro-photoluminescence spectroscopy. Its utility is proven by a systematic magnetooptical study of a single CdTe/ZnTe QD containing a Mn$^{2+}$ ion, where a magnetic field of up to 10 T in two orthogonal, Faraday and Voigt, configurations is applied to the same QD. The presented approach can be applied to a wide range of solid state nanoemitters.
\end{abstract}

%\keywords{CdTe, CdMnTe, CdTe:Mn, DMS, exciton}

% 75.50.Pp Magnetic semiconductors
% 75.30.Hx Magnetic impurity interactions
% 78.20.Ls Magnetooptical effects
% 71.35.Ji Excitons in magnetic fields; magnetoexcitons
%
\maketitle
%#################################### INTRODUCTION
%\section{Introduction}

Semiconductor quantum dots (QDs) hold great potential for valuable applications in modern optoelectronics. For instance, they act as optically\cite{Gerard:PRL1998, Vuckovic:APL2003} or electrically\cite{Yuan:Science2002} pumped, highly efficient \emph{on-demand} single-photon sources for possible applications in, e.g., quantum information processing schemes.\cite{Knill:Nature2000} As demonstrated recently,\cite{Besombes:PRL2004, Kudelski:PRL2007, Goryca:PRL2009, Besombes:PRL2009, Kobak:NatureCom2014} embedding a magnetic dopant (e.g., a Mn or Co ion) in the QD further enhances its potential for device implementations. Thanks to the \textit{s,p-d} exchange coupling between a magnetic ion and the carriers confined to the QD, an efficient all-optical manipulation\cite{Goryca:PRL2009, Besombes:PRL2009, Kobak:NatureCom2014} and readout\cite{Besombes:PRL2004, Kudelski:PRL2007, Goryca:PRL2009, Besombes:PRL2009, Kobak:NatureCom2014} of the ion's spin projection becomes possible, a milestone in the development of the emerging field of solotronics, that is electronics exploiting quantum properties of individual electrons, ions or defects.~\cite{Koenraad:NatureMat:2011, Kobak:NatureCom2014}

Since QDs are typically randomly distributed in a semiconductor matrix, permanent marking of their position is an indispensable prerequisite for fabrication of any kind of functional device, as well as for any systematic research involving QDs. Several approaches have been utilized so far for high accuracy positioning of individual solid-state nano-emitters, such as QDs or luminescent nanocrystals.\cite{Badolato:Science2005, Lee:APL2006Registr, Dousse:PRL2008, Thon:APL2009, vanderSar:APL2011, Rivoire:APL2011, Rabouw:OptMat2013, Gschrey:APL2013} The most efficient ones combined either \textit{in situ} optical lithography with micro-photoluminescence ($\mu$-PL)\cite{Dousse:PRL2008} or, offering even higher spatial resolution, \textit{in situ} electron beam lithography with cathodoluminescence.\cite{Gschrey:APL2013} The optical method (Ref.~\onlinecite{Dousse:PRL2008}) involved two laser beams of different wavelengths (\textit{two-color method}). The first beam served for determination of the position of the selected QD through $\mu$-PL mapping. Its energy was low enough to avoid exposure of a positive photoresist film deposited on the sample surface. The second laser beam, sharing the same optical path as the first one and of energy high enough to expose the photoresist, was switched on once the QD position was determined. As a result, a mark centered above the selected QD was obtained on the sample surface after the photoresist development. The method has proved its extraordinary efficiency for the production of deterministically coupled (In,Ga)As/GaAs QD - microcavity optical mode devices.\cite{Dousse:PRL2008, Dousse:APL2009, Dousse:Nature2010} The performance of the method is still limited, however, by the necessity of a precise alignment of the two laser beams and of an overlap of their focused spots on the sample surface.
\begin{figure}[b]
\vspace{-0.8cm}
\includegraphics[width=0.8\linewidth]{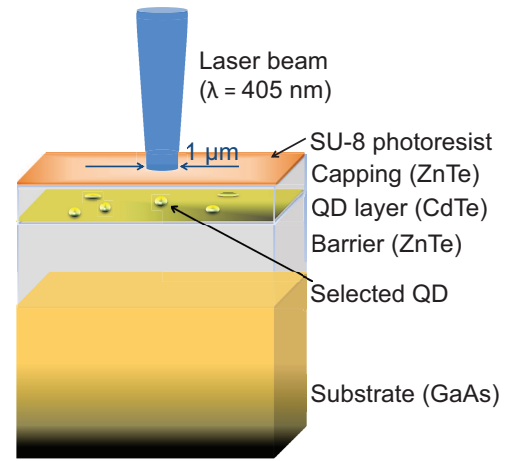}
\caption{Design of a sample embedding a Quantum Dot layer, with a photoresist spincoated onto its surface. A laser beam serves for determination of the QD position through PL-mapping and, after increase of the power density, for marking the QD position through exposure of the photoresist.
} \label{fig:sample3D}
\end{figure}
In this letter, we present a scalable method for the photolithography marking of the position of individual semiconductor QDs in a much simpler way than it has been performed so far.\cite{Badolato:Science2005, Lee:APL2006Registr, Dousse:PRL2008, Thon:APL2009} A single laser beam serves for both determination of the spatial position of the selected QD and for the exposure of a photoresist deposited on the sample surface (\textit{single-color method}) obtained through an increase of the excitation beam power density (see Fig.~\ref{fig:sample3D}). This makes our method similar to the cathodoluminescence one developed in Ref.~\onlinecite{Gschrey:APL2013}, where switching from the mapping to the marking stage was obtained by an increase of the electron dose. The proposed approach ensures that any possible inaccuracies resulting from an imperfect alignment of two laser beams of different colors\cite{Dousse:PRL2008} are avoided. In order to prove the utility of the presented method, we perform a systematic magneto-PL study on a selected CdTe/ZnTe QD with a single Mn$^{2+}$ ion. Marking the QD enables us to perform measurements in the Faraday configuration and, after rotating a single-axis magnet by 90~deg, to continue a study of the same QD in the Voigt configuration. With the present work, we extend the capabilities of QD position marking, so far limited solely to III-V semiconductors,\cite{Badolato:Science2005, Lee:APL2006Registr, Dousse:PRL2008, Thon:APL2009, vanderSar:APL2011, Rivoire:APL2011, Rabouw:OptMat2013} to a group of II-VI semiconductors. The II-VIs offer numerous advantages, like large exciton oscillator strength or negligible nuclear polarization of the host. The resulting enhanced light-matter coupling is profitable for the development of photonic and solotronic semiconductor devices operating at the ultimate, single photon level.~\cite{Pacuski:CrystGD2014, Jakubczyk:ACSNano2014}
%#################################################################
%\section{\label{sec:samples} Samples and experimental setup}

The samples containing self-assembled CdTe QDs embedded in a ZnTe barrier are grown by Molecular Beam Epitaxy on a GaAs substrate, as described in Ref.~\onlinecite{Gietka:APPA2012}. The QD layer with a planar QD density of $5\cdot10^9$/cm$^2$, buried 100~nm below the sample surface, is doped with a very low density of Mn$^{2+}$ ions. In the preparatory step, the sample is first cleaned in acetone and then in isopropyl alcohol in an ultrasonic bath in order to facilitate the subsequent process of photoresist deposition. It is then dried in a stream of pure nitrogen. A negative photoresist (MicroChem Nano SU-8 2002) is spin-coated (6100~rpm) onto the sample surface to sub-micrometer thickness.\cite{Lee:APL2006Registr} Next, the sample is baked at 100~$^\circ$C for 1~minute to evaporate the photoresist solvent.

For the photolithography marking of the QD position, the sample is placed inside a cold finger helium flow cryostat at a temperature $T$ = 8~K. The QD's emission is continuous-wave excited above the barrier bandgap at E$_{exc}$ = 3.06~eV ($\lambda_{exc}$ = 405~nm). The excitation beam is focused on the sample surface to a 1~$\mu$m diameter spot with a microscope objective (NA = 0.70, magnification = 100x, working distance = 6.5~mm) mounted on a piezo-electric X-Y-Z translation stage providing 20~nm precision of movement.

For the polarization resolved $\mu$-PL studies in a magnetic field of the marked QDs, the sample is placed in a pumped helium cryostat ($T$ down to 1.5 K) equipped with a superconducting split coil ($B$ up to $10$~T). The $\mu$-PL studies are conducted in Faraday or, after rotation of the cryostat by 90 deg, in Voigt configuration, that is with the magnetic field parallel or perpendicular to the direction of the light propagation, respectively. A lens of \textit{f} = 3.1~mm mounted on high accuracy piezo actuators focused the excitation beam to a 2~$\mu$m diameter spot on the sample surface. In both setups the signal is detected using a spectrometer with a CCD camera on its output (50~$\mu$eV of overall spectral setup resolution).
%#################################################################
%\section{\label{sec:litogr} \emph{Single-color} photolithography marking of individual Quantum Dots}

Photoluminescence studies of II-VI semiconductor based QDs, such as in the present work, typically necessitate a higher excitation energy than in the case of their III-V, e. g., GaAs based, counterparts. This is because of the much wider band gap of the materials used typically as barrier layers. Hence, in order to avoid photoresist exposure already at the  $\mu$-PL mapping\cite{Gschrey:APL2013} stage in the II-VI QDs case, a photoresist of correspondingly lower sensitivity must be applied. In our case, we chose the negative SU-8 2002 photoresist and used a laser beam of energy corresponding to the low energy tail of its sensitivity. The SU-8 2002 has the additional advantage of being practically fully transparent in the spectral region of the studied QD's emission. Also, it sustains well helium temperatures; in particular, no cracks are formed, unless the sample is cooled down from 300~K to 10~K within a time shorter than 0.5~h. We note that the flexibility of the proposed method enables, in principle, the application of a positive photoresist advantageous for, e.g., a lift-off process in further sample processing.

\begin{figure}
\includegraphics[width=1\linewidth]{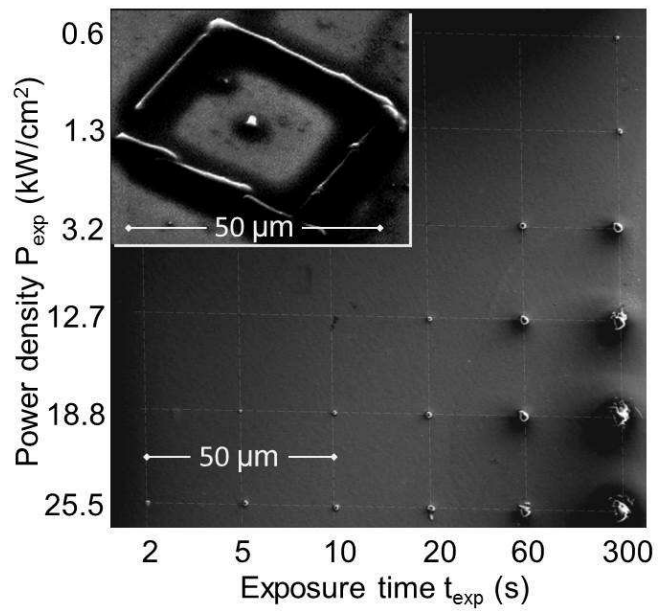}
\caption{Scanning Electron Microscope image of the sample surface with spots of developed Nano SU-8 2002 photoresist as a function of exposure time and of exposing beam power density. The lines are to guide the eye. Inset: A photoresist spot above the selected QD.} \label{fig:lithography}
\end{figure}
The Scanning Electron Microscope image shown in Fig.~\ref{fig:lithography} presents the result of a calibration of the SU-8 2002 photoresist deposited on the sample surface. A set of exposures is performed at $\lambda_{exc}$ = 405~nm with a power density $P_{exp}$ varied between 0.6~kW/cm$^2$ to 25.5~kW/cm$^2$ and the time $t_{exp}$ varied between 1~s to 300~s. The diameter of the smallest spot obtained after developing the photoresist is 1~$\mu$m, confirming the high spatial resolution of the setup. As expected for a Gaussian laser beam, the size of the exposed area increases with increasing $P_{exp}$ and/or $t_{exp}$. However, no traces of the exposed photoresist are evident for the lowest values of $P_{exp}$ and/or $t_{exp}$. This indicates that if the exposure dose absorbed by the photoresist does not exceed a threshold value, the exposure leaves the photoresist intact. In particular, this is the case for exposures with $t_{exp} \leq$~60~s and $P \leq$ 1.3~kW/cm$^2$. An individual QD emission spectrum is typically acquired in one second and with a power density of 0.01-0.1~kW/cm$^2$. This makes possible $\mu$-PL mapping with a small step (20~nm) without the risk of affecting the photoresist, as desired.

In the first stage of the actual QD position marking, the location of the selected QD is determined through X-Y, typically meander like, $\mu$-PL mapping with a minimum step of 20~nm over a surface of a few $\mu$m diameter. The mapping assures a precision in QD position determination of 50~nm, much less than the wavelength of the scanning laser, as in previous studies.\cite{Dousse:PRL2008, Thon:APL2009}
Once the QD position is determined and the laser beam is spotting the selected QD, the $P_{exp}$ is increased above 20~kW/cm$^2$ for $t_{exp}$ = 5 seconds in order to expose the photoresist. As a result, a circular spot on the sample surface located above the selected QD is obtained after the development process. The short exposure time eliminates any artifacts related to possible sample position drift. In order to facilitate identification of the marked spot in subsequent measurements, a marker in the form of a square (40~$\mu$m~$\times$~40~$\mu$m) is additionally exposed (see inset to Fig.~\ref{fig:lithography}). After marking the desired number of QDs, the sample is taken out of the cryostat and undergoes a post-exposure baking for 1~ minute at 100~$^\circ$C. Next, it is immersed for 1~minute in SU-8 developer at ambient temperature.
\begin{figure}
\includegraphics[width=0.8\linewidth]{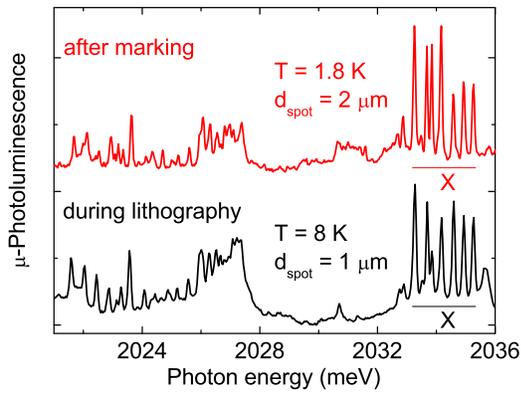}
\caption{Emission spectra of a selected CdTe/ZnTe QD containing a single Mn$^{2+}$ ion before (during the lithography step) and after the marking process acquired in two different experimental setups with a laser spot diameter of 1~$\mu$m and 2~$\mu$m, respectively. Excitation power density P$_{exc}$ = 0.02~kW/cm$^2$.} \label{fig:spectra}
\end{figure}
Fig.~\ref{fig:spectra} shows the emission spectra of a selected sample CdTe/ZnTe QD containing a single Mn$^{2+}$ ion before and after the position marking process, acquired in two different experimental setups. Despite the different diameters of the excitation laser spots (1~$\mu$m or 2~$\mu$m, respectively), it is clearly seen that the QD emission remains unaffected by the marking process. Since for each of sixteen marked QDs, the selected QD was found after the lithography step under the marker, we state that the process yield of the presented method is close to 100 \%. The alignment inaccuracy is much smaller than the diameter of the excitation laser spot (1~$\mu$m).

\begin{figure}
\includegraphics[width=1\linewidth]{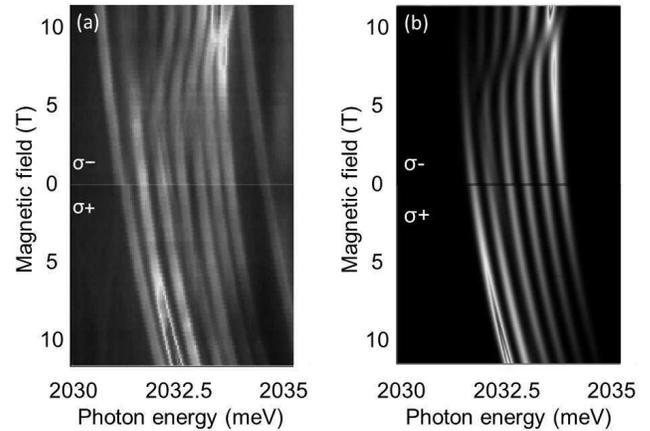}
\caption{a) Experimental and b) calculated emission spectra of a selected CdTe/ZnTe QD with a single Mn$^2+$ ion in a magnetic field applied in the Faraday configuration, for two circular polarizations of the light.}\label{fig:Faraday}
\end{figure}

In order to prove the utility of the presented QD position marking method, we perform a magneto-photoluminescence study of a neutral exciton (X) confined to the selected QD. In the absence of a magnetic field its emission spectrum consists of six lines centered at 2034~meV (see Fig.~\ref{fig:spectra}). Sixfold splitting originates from the exchange interaction between the anisotropic excitonic spin and the S = $5/2$ spin of the \textit{d}-shell electrons of the Mn$^{2+}$ ion.\cite{Besombes:PRL2004, Goryca:PRL2009, Kobak:NatureCom2014} Additional multiplets seen at lower energies originate either from a charged exciton or biexciton confined to the same QD or from excitons confined to neighboring QDs. Polarization dependent spectra in the Faraday configuration in magnetic fields up to 10~T are presented in Fig.~\ref{fig:Faraday}. As may be seen, the six emission lines shift in energy, according to the interplay of the Zeeman effect (linear dependence on magnetic field) and the diamagnetic shift (quadratic-like dependence). Around B = 8 T in $\sigma^-$ polarization anticrossings related to dark excitonic states are observed.\cite{Besombes:PRL2004, Goryca:PRB2010} The emission line seen at 2031.25~meV at B = 0~T shifts in a similar way to the low energy lines of the X. Its lack of polarization at higher magnetic field suggests, however, that it comes from a non-magnetic QD, neighboring the selected one.

%################################################################# Measurements in magnetic field
%\section{\label{sec:magnetoPL} Measurements}
\begin{figure}[t]
\includegraphics[width=1\linewidth]{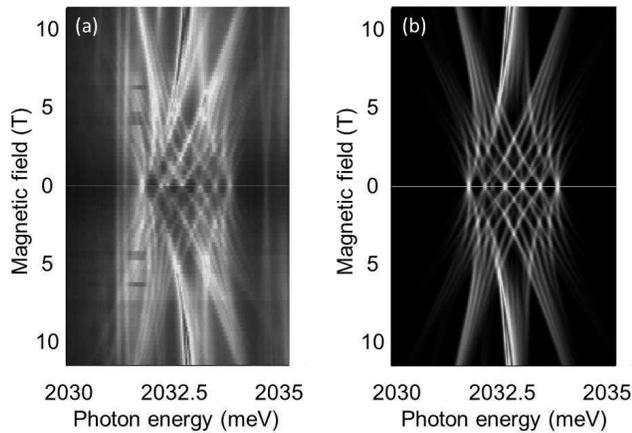}
\caption{The same as in Fig.~\ref{fig:Faraday}, but for the Voigt configuration and with no polarization resolution.} \label{fig:Voigt}
\end{figure}

Typically, the chance of finding a signal from the same QD after a change introduced to the experimental setup is very low. Here, thanks to QD marking, we are able to perform a measurement in the Voigt configuration on the same QD, after rotation of the cryostat by 90~deg. As seen in Fig.~\ref{fig:Voigt}, the X emission spectra in the Voigt configuration are very different from those measured in the Faraday configuration. As explained in detail in Ref.~\onlinecite{Leger:PRB2005} multiple splittings can be understood intuitively by noting that the influence of the exciton on the Mn$^{2+}$ ion spin state resembles that of an effective magnetic field approximately parallel to the QD growth axis. As a consequence, at low magnetic field the Mn$^{2+}$ spin is quantized along different axes in the presence or absence of the exciton. This leads to a sixfold splitting of each of the X emission lines. On the other hand, with increasing magnetic field the Mn$^{2+}$ spin becomes aligned and quantized along the field. In that case, the exchange interaction with the X introduces only a small admixture of other ion spin states to the field-induced one. This results in a strengthening of the middle group of X emission lines (associated with no Mn$^{2+}$ spin change during the X recombination), as well as in a weakening of lateral groups of lines (recombination with a change of the ion spin). The negligible shift of the emission lines seen at around 2031.25~meV at B = 0~T provides an unequivocal indication that they come from a non-magnetic QD.

The emission spectra of X confined in a CdTe/ZnTe QD containing a single Mn$^{2+}$ ion presented in Figs.~\ref{fig:Faraday}b and \ref{fig:Voigt}b are calculated following the Hamiltonian $H = H_B + H_{exch} + H_{lh-hh}$.~\cite{Besombes:PRL2004,Leger:PRB2005,Goryca:PRB2010} The $H_B = g_{Mn}\mu_B\vec{S}\vec{B} + g_{e}\mu_B\vec{\sigma}\vec{B} + g_{h}\mu_B\vec{j}\vec{B} + \gamma B_z^2$ term describes the interaction of the system with the magnetic field;  the $H_{exch} = - \frac{2}{3}\delta_0 \sigma_z j_z + \frac{2}{3} \delta_1 (\sigma_y j_y^3 - \sigma_x j_x^3) + I_e\vec{\sigma}\vec{S} + I_h \vec{j}\vec{S}$ term, the exchange interaction between carriers; and the $H_{lh-hh} = \Delta(j_x^2+j_y^2) - \delta(j_x^2-j_y^2)$ term, the heavy-light hole splitting and mixing. Factors $S$, $\sigma$ and $j$ denote the Mn$^{2+}$, electron and hole spin operators, respectively. The best agreement between the model and the experiment is obtained for $g$-factors $g_e=-0.4$ and $g_h=0.43$, diamagnetic constant $\gamma=2.9\cdot10^{-3}$~meV/T$^2$, isotropic (anisotropic) electron-hole exchange constant $\delta_0=0.65$ meV ($\delta_1=0$~meV), and electron- and hole-Mn$^{2+}$ exchange constants $I_e=-0.085$~meV and $I_h=0.24$~meV, respectively. Heavy-light hole splitting and mixing parameters are taken to be $\Delta=30$ meV and $\delta=2.2$ meV. As is seen, the modeling provides reasonable values of the characteristic constants.\cite{Besombes:PRL2004, Leger:PRB2005, Goryca:PRB2010, Kobak:NatureCom2014}
To reproduce the relative intensities of the exciton emission lines an effective temperature $T=20$~K of the Mn$^{2+}$ ion is introduced.\cite{Besombes:PRL2004, Leger:PRB2005, Goryca:PRB2010, Kobak:NatureCom2014}
%#################################################################
%\section{Conclusions}

In summary, a simple, scalable method for the permanent marking of the position of individual II-VI QDs is demonstrated. The \emph{single-color} marking process involves \textit{in situ} photolithography combined with low-temperature $\mu$-PL. The emission spectra shown provide a direct comparison of the magnetic field dependence of a CdTe/ZnTe QD with and without a magnetic ion, indicating a method for unequivocal identification of magnetically doped QD emission lines. We have checked that the markers act as a durable protection of the sample against etching with a beam of gallium ions with standard current and acceleration voltage values. It provides a chance for the preparation of mesa structures like, e. g., micropillars,\cite{Jakubczyk:APL2012} containing the selected QD using of the markers as protection masks, similar to what was done in Ref.~\onlinecite{Gschrey:APL2013}. Thanks to its versatility, the photolithographic technique presented here could be applied to a wide range of nanoemitters, like N-V centers in diamond or colloidal QDs, facilitating their advanced implementations, e.g., in quantum communication schemes involving networks of distant emitters coupled through an optical cavity mode.

%\section*{Acknowledgments}

This work was supported by the Polish National Research Center projects DEC-2011/02/A/ST3/00131 and NCN 2013/10/E/ST3/00215, and by the Polish National Center for Research and Development project LIDER.

\end{document}